\documentclass[letterpaper,12pt,preprint]{emulateapj}

\def\solmass {\,$\hbox{M}_\odot$}

\def\nthp {N$_2$H$^+$}

\shorttitle{Substructure In Starless Cores}
\shortauthors{Schnee et al.}

\begin{document}

\title{An Observed Lack of Substructure in Starless Cores} 

\author{Scott Schnee\altaffilmark{1}, Melissa Enoch\altaffilmark{2},
Doug Johnstone\altaffilmark{1,3}, Thomas Culverhouse\altaffilmark{4,5},
Erik Leitch\altaffilmark{4,5}, Daniel P Marrone\altaffilmark{4,5},
Anneila Sargent\altaffilmark{6}}

\email{scott.schnee@nrc-cnrc.gc.ca}

\altaffiltext{1}{National Research Council Canada, Herzberg Institute
of Astrophysics, 5071 West Saanich Road Victoria, BC V9E 2E7, Canada}
\altaffiltext{2}{Department of Astronomy, University of California,
Berkeley, CA 94720, USA}
\altaffiltext{3}{Department of Physics \& Astronomy, University of
Victoria, Victoria, BC, V8P 1A1, Canada}
\altaffiltext{4}{Kavli Institute for Cosmological Physics, Department
of Astronomy and Astrophysics, University of Chicago, Chicago, IL
60637, USA}
\altaffiltext{5}{Department of Astronomy and Astrophysics, University
of Chicago, Chicago, IL 60637, USA}
\altaffiltext{6}{Department of Astronomy, California Institute of
Technology, MC 105-24 Pasadena, CA 91125, USA}

\begin{abstract}

In this paper we present the results of a high resolution (5\arcsec)
CARMA and SZA survey of the 3\,mm continuum emission from 11 of the
brightest (at 1.1\,mm) starless cores in the Perseus molecular cloud.
We detect 2 of the 11 cores, both of which are composed of single
structures, and the median 3$\sigma$ upper limit for the
non-detections is 0.2\solmass in a $\sim$5\arcsec\ beam.  These
results are consisent with, and as stringent as, the low detection
rate of compact 3\,mm continuum emission in dense cores in Perseus
reported by \citet{Olmi05}.  From the non-detection of multiple
components in any of the eleven cores we conclude that starless core
mass functions derived from bolometer maps at resolutions from
10\arcsec--30\arcsec\ (e.g.~with MAMBO, SCUBA or Bolocam) are unlikely
to be significantly biased by the blending of lower mass cores with
small separations.  These observations provide additional evidence
that the majority of starless cores in Perseus have inner density
profiles shallower than $r^{-2}$.

\end{abstract}


\section{Introduction}

Surveys of starless cores in nearby molecular clouds have shown that
the starless core mass function (CMF) looks very similar to the
stellar initial mass function (IMF), but translated towards higher
masses by a factor of a few \citep[e.g.][]{Motte98, Nutter07, Enoch08,
Simpson08}.  If each starless core collapses into a single star, then
the apparent shift between the IMF and the CMF implies that the
formation process is inefficient and most of the mass in each core
never makes it into any star.  Models have shown that 25\%-70\% of the
mass in a protostellar core will be converted into a star with the
remaining material being ejected by outflows \citep{Matzner00}.  A
second reason for the shift between the CMF and the IMF is likely to
be fragmentation, in which a single starless core will eventually form
multiple stars.  Since many stars are found in binary or higher-order
systems \citep[e.g.][]{Duquennoy91}, to some extent the fragmentation
of a starless core into multiple bound objects is to be
expected. However, lower mass stars are less frequently found in
gravitationally bound systems \citep{Lada06}.

Although the shape of the IMF can be reproduced from the CMF with a
variety of assumptions about the multiplicity, fragmentation, and
efficiency of the process \citep{Swift08, Hachell08}, recent work by
\citet{Goodwin08} has shown that models in which every starless core
eventually forms multiple stars reproduces the stellar IMF more
successfully than models in which low-mass cores are more likely to
form a single star than a binary.  That each core fragments into only
two or three components, and not many more, is suggested by the
relatively small number of close binaries and high velocity single
stars that would result from the dynamical ejection of protostars from
higher order systems \citep{Goodwin05}.  Therefore, high angular
resolution observations of starless cores might be expected to show
that what appears to be a single object at low resolution is in fact
composed of multiple fragments, depending on whether or not the
fragmentation takes place before an embedded protostar forms and on
what physical scale fragmentation occurs.

The binary fraction of young stellar objects and protostars has been
studied in detail, and it has been found that these objects are
commonly found in multiple systems.  For example, near-infrared
observations of Class-I and flat-spectrum objects in nearby molecular
clouds show that in the range 300-2000\,AU the multiplicity is
$\sim$18\%, with a median separation between companions of
$\sim$900\,AU \citep{Haisch04}.  Sub-arcsecond
Berkeley-Illinois-Maryland Array (BIMA) observations of the 3\,mm
continuum emission from 20 Class 0 and Class I protostars found that
all of the embedded objects are in small groupings, with separations
ranging from $<$100\,AU to $>$10000\,AU \citep{Looney00}.  Recent
simulations of fragmentation in magnetically supported dense cores
typically start with small cores ($\sim$6000\,AU or 24\arcsec\ at the
distance of Perseus) that fragment into multiple protostellar systems
with separations on the order of 10--1000\,AU
\citep[e.g.][]{Boss02,Boss07,Price07}.

Observations of starless cores have not often found evidence for
fragmentation on the scale of 5--10\arcsec.  In a survey of dense
cores in the Perseus molecular cloud with the Owens Valley Radio
Observatory (OVRO) interferometer, \citet{Olmi05} found that five out
of six cores had no compact continuum emission, and in the sixth core
a single source was detected (at $\sim$5\arcsec\ resolution, or
1200\,AU in Perseus).  Similarly, \citet{Harvey03} mapped the 1.3\,mm
continuum emission from the starless core L694-2 with the IRAM Plateau
de Bure Interferometer (PdBI) and BIMA and found that the PdBI
observations resolved out the core entirely while the shorter
baselines of the BIMA data can be adequately modelled by a single
source, implying that no fragmentation has taken place.  Other
interferometric observations of starless cores, such as S68NW, L1544
and L694-2 show only single peaks in their integrated spectral line
emission \citep{Williams99a, Williams99b, Williams06}.  However, BIMA
mapping of the \nthp\ emission from the starless core L183 shows that
this core is composed of three components with a total mass of a few
tenths of a solar mass that are distinct both spatially and
kinematically \citep{Kirk09}.  \citet{Kirk09} interpret this as
evidence that L183 is fragmenting as it collapses and predict that
this core will eventually form multiple protostars.

Here we present the results of a search for substructure in a sample
of 11 starless cores in the Perseus molecular cloud.  We use Combined
Array for Research in Millimeter-wave Astronomy (CARMA) and
Sunyaev-Zel'dovich Array (SZA) interferometric observations to study
the 3\,mm continuum emission at $\sim$5\arcsec\ resolution with
sensivity up to scales of 30\arcsec - 80\arcsec.  The source selection
and observations are discussed in Section \ref{OBSERVATIONS}, followed
by an analysis (Section \ref{ANALYSIS}), a discussion (Section
\ref{DISCUSSION}), and summary (Section \ref{SUMMARY}).

\section{Observations} \label{OBSERVATIONS}

To search for substructure within starless cores, we mapped the 3\,mm
continuum emission with $\sim$5\arcsec\ resolution towards a sample of 
11 cores drawn from the 1.1\,mm Bolocam survey of \citet{Enoch08}.  Here 
we describe the sample of cores and their observations.

\subsection{Source Selection} \label{SELECTION}

We drew our sample of starless cores from the 1.1\,mm Bolocam survey
of \citet{Enoch06, Enoch08}, which has 31\arcsec\ resolution.
\citet{Enoch08} uses {\it Spitzer} near- to mid-infrared data to
determine the starless/protostellar status of each millimeter core.
We chose starless cores in the Perseus molecular cloud on the basis of
having high peak fluxes ($>$200 mJy beam$^{-1}$).  There are 15
starless cores in the Bolocam survey of Perseus with peak fluxes
$>$200 mJy beam$^{-1}$, of which we were only allocated enough time to
observe 11, chosen arbitrarily.  The selected cores have a median mass
of 1.47\solmass\ and a median density of $3.2 \times 10^5$ cm$^{-3}$,
as derived from the Bolocam data \citep{Enoch08}.  The 1.1\,mm-derived
properties of the starless cores in this sample are presented in Table
\ref{BOLOCAMTAB}.

Protostars in \citet{Enoch08} are identified by the shape of their
near-infrared to far-infrared spectral energy distributions (SEDs),
minimum flux at 24\,\micron\ and presence of 70\,\micron\ point
sources not classified as galaxy candidates.  In \citet{Enoch08}, if a
protostar is found within one FWHM of the center of the 1.1\,mm
Bolocam core, then it is classified as protostellar, otherwise the
Bolocam core is classified as starless.  Searches for protostars in
the Perseus molecular cloud have also been performed by
\citet{Jorgensen07}, \citet{Hatchell07}, and \citet{Sadavoy10}.
\citet{Jorgensen07} identify protostellar cores from their
mid-infrared colors and magnitudes, distance to the nearest
850\,\micron\ SCUBA core and concentration of the 850\,\micron\
emission.  \citet{Hatchell07} identify protostellar cores by the
presence of $^{12}$CO outflows and the presence of near-infrared
(2MASS) or mid-infrared ({\it Spitzer}) sources within an
850\,\micron\ SCUBA core.  \citet{Sadavoy10} find protostars based on
mid-infrared and far-infrared ({\it Spitzer}) magnitudes and colors
cross-referenced with a list of 850\,\micron\ SCUBA cores.  A detailed
comparison between these methods of classifying a core as either
starless or protostellar is given in \citet{Sadavoy10}; here we note
that the classifications according to the various criteria are broadly
consistent.

All of the cores in our sample are classified as starless by
\citet{Enoch08} and none of these cores are identified as protostellar
by \citet{Jorgensen07}, but Perbo45 and Perbo50 are identified as
Class 0 protostars by \citet{Hatchell07} and Perbo44 and Perbo107 are
identified as protostars by \citet{Sadavoy10}.  More recent deep
70\,\micron\ observations of one starless core in our sample, Perbo58,
indicate that it may contain a very weak internal luminosity source
(Enoch et al., in prep).  Thus, Perbo58 may not be truly prestellar,
although even if a central protostar has formed, it is likely very
young and not far evolved from the prestellar phase.  We keep Perbo44,
Perbo45, Perbo50, Perbo58 and Perbo107 in the starless sample based on
the original selection criteria and note that out of the four surveys
to identify starless cores and/or protostars in the Perseus molecular
cloud none of our targets are classified as protostellar in more than
one paper.

\subsection{CARMA and SZA} 

Continuum observations in the 3\,mm window were obtained with CARMA, a
15 element interferometer consisting of nine 6.1 meter antennas and
six 10.4 meter antennas.  The CARMA correlator records signals in
three separate bands, each with an upper and lower sideband.  We
configured all three bands for maximum bandwidth (468\,MHz with 15
channels per band) to observe continuum emission, providing a total
continuum bandwidth of 2.8\,GHz.  The observations were centered around
102\,GHz, and range from 100\,GHz to 104\,GHz.  The half-power beam width
of the 10.4\,m antennas is 66\arcsec\ at the observed frequencies.
Single pointings towards all 11 starless cores were taken in the CARMA
D-array configuration, with projected baselines that range from 11\,m
to 150\,m.  Follow-up observations of six of the eleven cores were made
in the CARMA E-array cofiguration with projected baselines that range
from 8\,m to 66\,m, also with single pointings.  Two cores (Perbo45 and
Perbo58) were then re-observed in D and E-array configurations with
7-point mosaics to improve the signal-to-noise in the maps and enlarge
the area of uniform gain to 66\arcsec.  A summary of the CARMA
configurations, the synthesized beam size, the largest angular scale
to which the observations are sensitive, and the noise in the maps are
presented in Table \ref{MMOBSTAB}.

The SZA is an 8 element interferometer equipped with 3\,mm band
(80--115\,GHz) receivers.  The instantaneous field of view of the SZA
is given by the primary beam of the 3.5\,m antennas, approximately
3.3\arcmin\ (FWHM) at 95\,GHz.  Six of the SZA antennas are arranged
in a close-packed configuration with baselines ranging from 4.5 to
11.5\,m and two outer telescopes yield baselines of up to 65\,m,
resulting in a synthesized beam size of 38\arcsec.  Sixteen 500\,MHz
wide analog bands provide 8\,GHz bandwidth, though for the
observations in this study only 7\,GHz bandwidth (from 92--98\,GHz)
was available.  A more detailed description of the SZA is given in
\citet{Muchovej07}.

The observing sequence for the CARMA observations was to integrate on
a phase calibrator (3C84, 0237+288 or 0336+323) for 4--5 minutes and
science targets for 14--28 minutes.  In each set of observations 3C84
was observed for passband calibration, and observations of a planet
(Mars or Uranus) or 3C84 were used for absolute flux calibration.
Although 3C84 has a variable brightness, it is monitored on a weekly
basis at CARMA.  Based on the repeatability of the quasar fluxes, the
estimated random uncertainty in the measured source fluxes is
$\sigma\sim5$\%.  Radio pointing was done at the beginning of each
track and pointing constants were updated at least every two hours
thereafter, using either radio or optical pointing routines.
Calibration and imaging were done using the MIRIAD data reduction
package \citep{Sault95}.

The observing sequence for the SZA observations was to integrate on
the phase calibrator, 3C84, for 5 minutes and the starless core
Perbo58 for 10 minutes.  The observations of 3C84 were also used for
bandpass and absolute flux calibration.  Because the SZA data are at a
somewhat lower frequency than the CARMA data (95\,GHz vs.~102\,GHz),
we use the flux of 3C84 derived at the higher frequency to set the
absolute flux calibration of the SZA data before combining the CARMA
and SZA maps.  Calibration and imaging were done using the MIRIAD data
reduction package \citep{Sault95}.

\begin{figure}
\epsscale{0.50} 
\plotone{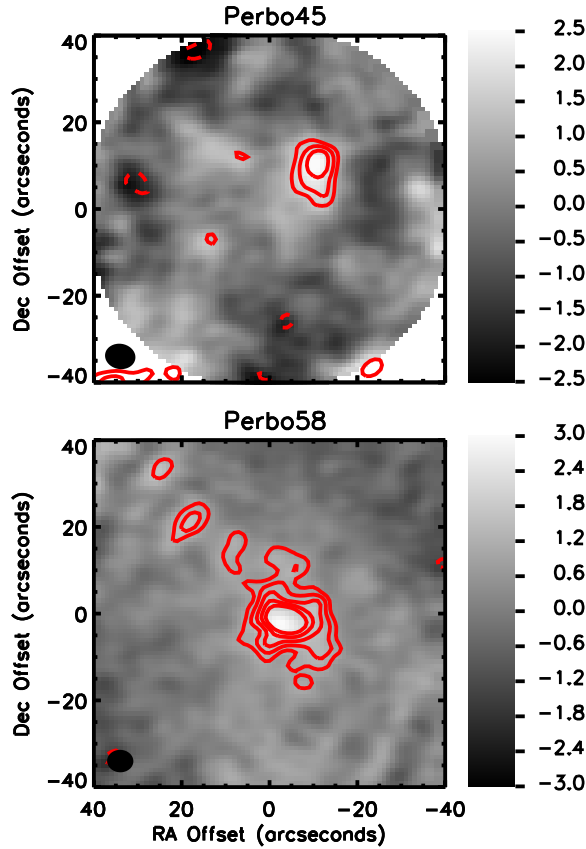}
\caption{3\,mm flux maps (grayscale and contours) of Perbo45 (top)and
Perbo58 (bottom), from data taken with CARMA and, for Perbo58, SZA.
Contours start at 3$\sigma$ and increase by 1$\sigma$, negative peaks
are shown with dashed contours. The beam size is shown in the bottom
left corner of each map.  The units of the maps are mJy beam$^{-1}$.
\label{CARMASZAMAPS}}
\end{figure}

The 3\,mm continuum maps of Perbo45 and Perbo58 are presented in
Figure \ref{CARMASZAMAPS}.  The other starless cores were not
detected.

\subsection{Bolometer Maps} \label{BOLOMETER}

\begin{figure}
\epsscale{1.00} 
\plotone{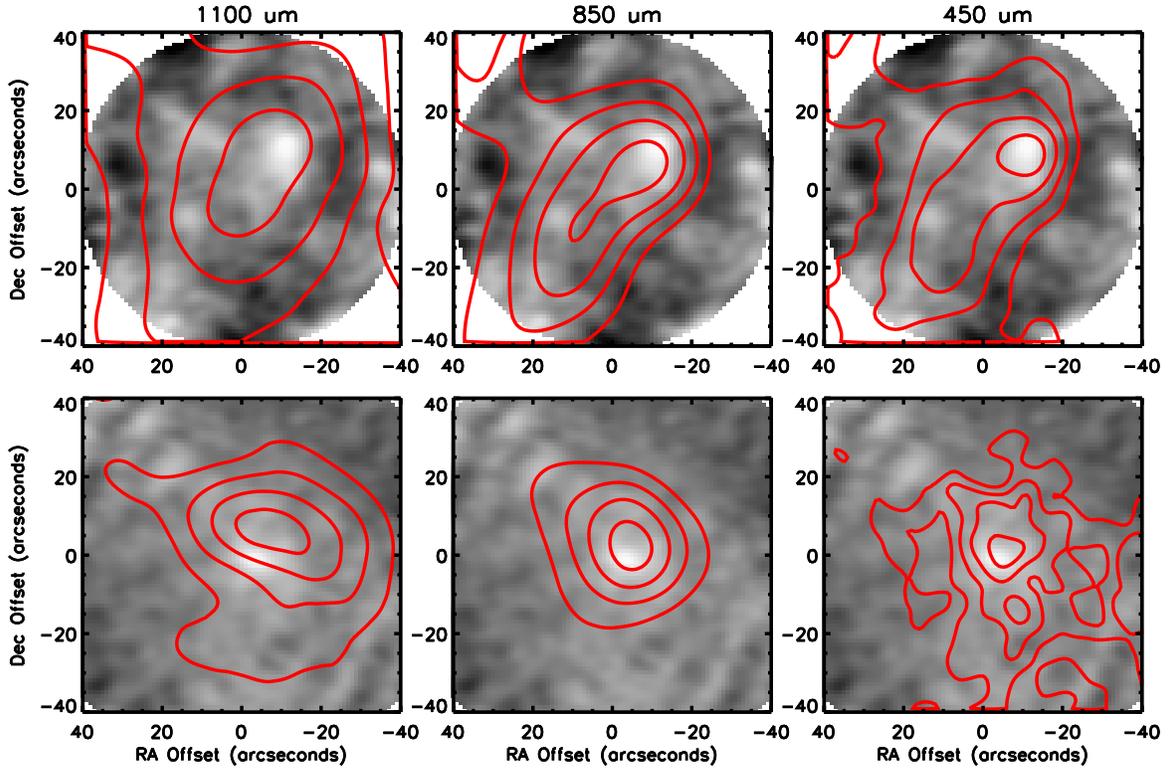}
\caption{3\,mm flux maps (grayscale) of Perbo45 (top) and Perbo58
(bottom), with the same scale as in Figure \ref{CARMASZAMAPS}.
Contours show the emission from maps made by Bolocam 1.1\,mm (left),
SCUBA 850\,\micron\ (middle) and SCUBA 450\,\micron\ (right), with
contour levels at 30\%, 50\%, 70\%, and 90\% of the peak in each map.
\label{OVERLAYMAPS}}
\end{figure}

In addition to the 31\arcsec\ Bolocam map at 1100\,\micron\
\citep{Enoch06}, the Perseus molecular cloud has also been mapped with
SCUBA at 450 and 850\,\micron\ at 9\arcsec\ and 14\arcsec\ resolution,
respectively \citep{Hatchell05, Kirk06}.  We downloaded the
submillimeter maps from the SCUBA Legacy Catalogues
\citep{diFrancesco08}, which includes all of the data taken with SCUBA
and uses the most current calibration and reduction methods.  The
absolute calibration of the bolometer maps is good to 15\%, 20\% and
50\% at 1100, 850 and 450\,\micron, respectively.  The 450, 850 and
1100\,\micron\ bolometer maps of Perbo45 and Perbo58 are overlayed on
the 3\,mm interferometric maps in Figure \ref{OVERLAYMAPS}.

\section{Analysis} \label{ANALYSIS}

We detected, at the $\ge$5$\sigma$ level, continuum emission from 2
out of 11 starless cores in our survey.  Here we present an analysis
of the properties of the detected cores and the implications of this
detection rate.

\subsection{Core Properties} \label{PROPERTIES}

Both of the detected cores in this survey (Perbo45 and Perbo58) are
seen to contain one main component (see Figure \ref{CARMASZAMAPS}).  A
few lower significance (3-4$\sigma$) peaks are present in the 3\,mm
map of Perbo58 along an extended ridge seen in the bolometer maps.
Note that the map of Perbo58 includes data taken with the SZA, so it
is sensitive to angular scales smaller than $\sim$80\arcsec.  Given
that the FWHM sizes of Perbo58 as measured by Bolocam (26\arcsec) is
smaller than this scale, we believe that we have not resolved out any
of the 3\,mm emission from Perbo58, but given that the map of Perbo45
only includes CARMA D+E array data and not SZA data we are likely
resolving out some of the 3\,mm emission from Perbo45.

\subsubsection{Mass} \label{MASS}

We calculate the mass of the detected cores from the total 3\,mm flux
and upper limits to the non-detections using the equation:
\begin{equation} \label{MASSEQ}
M = \frac{d^2 S_{\rm 3mm}}{B_\nu(T_D)\kappa_{\rm 3mm}}
\end{equation}
where $d$ is the distance to Perseus, $S_{\rm 3mm}$ is the total flux
from each detection or three times the noise for the non-detections,
$B_\nu$ is the Planck function at dust temperature $T_D$, and
$\kappa_{\rm 3mm} = 0.00169$ cm$^2$g$^{-1}$ is the dust opacity,
extrapolated from the opacity at 1300\,\micron\ given in
\citet{Ossenkopf94} Table 1, column (5) for dust grains with thin ice
mantles, assuming a gas-to-dust ratio of 100 and an emissivity given
by $\beta = 2$.  This value of the emissivity spectral index is
consistent with the measured $\beta$ in TMC-1C \citep{Schnee10} and
L1498 \citep{Shirley05}.  For comparison with the core properties
derived in \citet{Enoch08}, we assume that the distance to Perseus is
250\,pc and that the dust temperature is 10\,K for all of the cores in
this sample, which is a typical gas temperature of starless cores in
Perseus \citep{Schnee09}.  The derived masses and upper limits are
presented in Table \ref{MMFITSTAB}.

The mass of Perbo45 derived from our 3\,mm data is about a factor of 4
lower than the mass derived from the 1.1\,mm Bolocam data.  The mass
of Perbo58 derived from our 3\,mm observations is a factor of 3 higher
than the mass derived from the 1.1\,mm Bolocam data.  Four
possibilities for explaining these discrepancies are 1) uncertainties
in the dust properties 2) variations in the temperature within the
cores, 3) errors in the measurement of the 1.1\,mm and/or 3\,mm
fluxes, and 4) in the case of Perbo45, we are likely to have resolved
out some of the 3\,mm continuum emission.  Note that the deconvolved
size of Perbo45 in the Bolocam surveys is roughly 53\arcsec, so the
amount of flux resolved out by our CARMA observations is likely to be
substantial.

The masses of Perbo58 derived from the 1.1\,mm and 3\,mm data could be
brought into agreement if the emissivity spectral index were given by
$\beta \simeq 0.5-1$ instead of $\beta = 2$, as we assumed.  Although
the two best measurements of the emissivity spectral index in starless
cores suggest that $\beta \geq 2$ is a much better fit than $\beta
\leq 1$ \citep{Shirley05, Schnee10}, it is possible that grain growth
leads to a shallower emissivity spectral index at the high densities
likely to be found at the centers of prestellar cores.  For instance,
the emissivity spectral index in a sample of nearby protostars has
been measured to be in the range $0.25-1.5$ \citep{Arce06}, and values
in this range might be plausible for starless cores given the recent
evidence for grain growth in starless cores \citep{Steinacker09}.
Because the masses in Table \ref{MMFITSTAB} are derived from the
1.3\,mm opacity \citep{Ossenkopf94} extrapolated out to 3\,mm, an
incorrect assumption about the emissivity spectral index would lead to
masses off by a factor of a few.  If the emissivity spectral index in
starless cores is closer to $\beta = 1$ than $\beta = 2$, then the
expected fluxes at 3\,mm as extrapolated from the 1.1\,mm Bolocam
observations would be a factor of $\sim$3 larger than we have assumed.
In this case the non-detection of 3\,mm continuum emission towards
nine of the eleven cores in our sample becomes an even more
significant result (see Section \ref{RATE}).

Alternatively, if Perbo58 were significantly warmer at the center than
its average temperature, the higher resolution 3\,mm observations
would be more biased towards high masses and densities than the low
resolution 1.1\,mm data.  It is generally observed that starless cores
are colder at their centers due to self-shielding
\citep[e.g.][]{Crapsi07, Schnee07}.  Heating due to contraction can
result in a temperatures a few degrees higher in the center of a
starless core \citep{Keto09}, but this on its own cannot account for
the large difference in masses derived from the 1.1\,mm and 3\,mm
continuum data.  If there were a weak internal luminosity source in
Perbo58 (as discussed in Section \ref{SELECTION}), then one would
expect some central heating that could bring the 1.1\,mm and
3\,mm-derived masses into agreement.

A third possible cause for the disagreement between the 1.1\,mm and
3\,mm derived masses is the uncertain absolute calibration of the
emission maps.  The two cores detected by CARMA/SZA are both in the
dense and bright cluster NGC1333, which results in artifacts in the
bolometer maps.  However, detailed testing of the accuracy of the
recovery of 1.1\,mm flux from Bolocam has shown that the quoted fluxes
are accurate to within 15\% \citep{Enoch06}.  The absolute calibration
of the 3\,mm continuum maps is accurate to within 20\%.  Therefore,
uncertainties in the absolute calibration of the emission maps alone
cannot account for the factor of a few difference between the 1.1\,mm
and 3\,mm derived masses.

Due to uncertainties in the dust emissivity and temperature profiles
of Perbo45 and Perbo58, we will defer a more detailed analysis of
their masses until more data are available and only note that the
3\,mm fluxes of Perbo45 and Perbo58 are not what we expected based on
their 1.1\,mm fluxes as mapped by Bolocam.

\subsubsection{Density} \label{DENSITY}

We calculate the density of the detected starless cores from the
derived mass and effective radius using the equation:
\begin{equation} \label{DENSITYEQ}
n = \frac{3M}{4 \pi r_{\rm eff}^3 \mu m_H}
\end{equation}
where $r_{\rm eff}$ is the effective radius, $\mu = 2.33$ is the mean
molecular weight per particle and $m_H$ is the mass of hydrogen.  The
effective radius is the geometric mean of the deconvolved semi-major
and semi-minor axes, derived from a Gaussian fit to the flux
distribution.  The density for each detected core is presented in
Table \ref{MMFITSTAB}.

The densities of Perbo45 and Perbo 58 derived from the 3\,mm continuum
data in this survey are larger than the values derived from the
1.1\,mm Bolocam data.  The densities derived from the 3\,mm data (see
Table \ref{MMFITSTAB}) are roughly a few $\times 10^6 - 10^7$
cm$^{-3}$, while the mean density in a $10^4$\,AU diameter aperature
derived from the 1.1\,mm data are on the order of a few $\times 10^5$
cm$^{-3}$.  Since the 3\,mm observations are higher resolution than
the 1.1\,mm observations, it is not surprising that the 3\,mm-derived
densities are larger.  Though high, densities in the $10^6 - 10^7$
cm$^{-3}$ range have been reported in L1544 \citep{Crapsi07} and in
several starless cores in Orion \citep{Nutter07}.  The densities in
Table \ref{MMFITSTAB} would be lower if the assumed temperature were
higher, as might be expected if the detected cores harbor
low-luminosity protostars.

A simple model of the density profile of a starless core can be given
by:
\begin{equation} \label{MODELEQ}
n(r) = \frac{n_0}{1+\left(\frac{r}{r_0}\right)^\alpha}
\end{equation}
where $n_0$, $r_0$ and $\alpha$ are constants.  This formulation with
$\alpha=2.5$ is a good approximation for the density profile of a
Bonnor-Ebert sphere \citep{Tafalla04} and provides a qualitatively
good fit to the observed density profiles of starless cores, which are
fairly flat at small radii and are steeper at larger radii.

\begin{figure}
\epsscale{1.00} 
\plotone{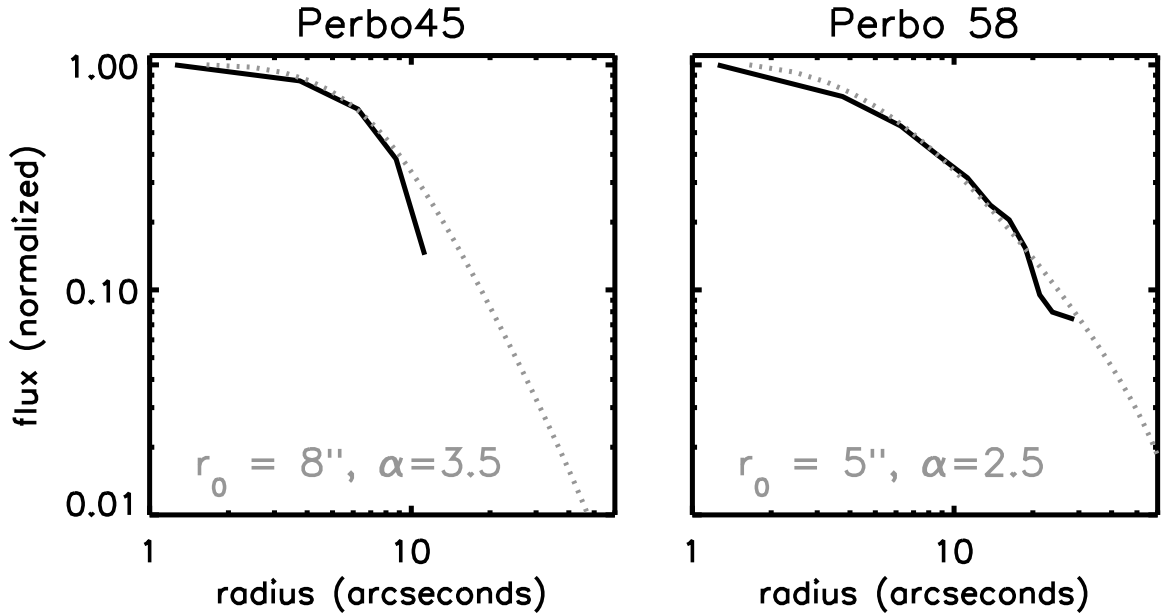}
\caption{3\,mm flux profiles of Perbo45 (left) and Perbo58 (right),
shown in solid black lines.  The flux profiles of model cores with
density profiles given by Equation \ref{MODELEQ} are shown in dotted
grey lines.  The inner point on the observed mass profiles is placed
at 1.25\arcsec, which is half of the HWHM of the typical beam size of
the 3\,mm observations.  Values of $r_0$ and $\alpha$, as described in
Section \ref{DENSITY}, are given in each panel.
\label{DENSITYMAPS}}
\end{figure}

We investigate whether or not the density profiles of the cores in
Perbo45 and Perbo58 are consistent with Equation \ref{MODELEQ}.  First
we measure the flux profile in annuli centered on the positions of
Perbo45 and Perbo58 given in Table \ref{MMFITSTAB}.  We then calculate
the flux profile that would be observed for a spherical core with the
density profile given by Equation \ref{MODELEQ} assuming that the
cores are isothermal and have constant dust emissivities.  We
normalize the observed and model fluxes such that the flux at the
center of the core is equal to unity, so the $n_0$ term in Equation
\ref{MODELEQ} is not meaningful.  We show in Figure \ref{DENSITYMAPS}
that under the stated assumptions of the dust properties and
geometries that a density profile like that given in Equation
\ref{MODELEQ} does predict flux profiles that are similar to those
observed at 3\,mm for values of $r_0$ in the range of
5\arcsec--8\arcsec and $\alpha$ in the range 2.5--3.5.

The model density profile that we use to fit the observed flux
profiles of Perbo45 and Perbo58 does not provide a unique solution to
what the true density profiles of these core are, and other models
that have a flat inner profile that steepens at larger radii can be
consistent with our observations.  If one uses different assumptions
about the temperature profile and allows the emissivity to vary with
radius, it might even be possible to fit the 3\,mm flux profile with a
power-law density profile.  What we have shown here is that the
density profiles of Perbo45 and Perbo58 are not necessarily different
from typical starless cores, and the reason that they are detected in
our survey while other cores are not is most likely due to their
relatively large masses for their size.

\subsection{Detection Rate} \label{RATE}

Of the 11 starless cores that were observed with CARMA, 2 were
detected (Perbo45 and Perbo58) and 9 were not.  Neither of the
detected starless cores are seen to break into multiple components.
The non-detection of the majority of the starless cores is not due to
low sensitivity (the median upper limit for the non-detections is
0.2\solmass\ - see Section \ref{MASS} for details), but instead is
likely the result of a lack of compact sub-structures within the
cores.  If all of the mass in each of the non-detected cores had been
contained within a few compact structures, every core would have been
detected in the CARMA observations (see Section \ref{MASS} for an
explanation of how the mass upper limits in Table \ref{MMFITSTAB} were
derived).  For example, the median 3$\sigma$ upper limit on the mass
of the non-detected cores is a factor of 7 smaller than the core mass
derived from the Bolocam map.  Therefore, we argue that the
non-detection of multiple components in all 11 cores in our sample is
a significant result.  Furthermore, given the high rate of
multiplicity in Class 0 and Class I protostars \citep{Looney00} and
the low rate of multiplicity in starless cores, we suggest that
fragmentation into binaries takes place during the collapse of a
prestellar core or during the Class 0 stage.

The density profiles of many starless cores have been measured from
(sub)millimeter continuum emission maps and it has generally been
found that the density profiles are consistent with an $r^{-2}$
power-law at large radii that flattens towards the center of the core
\citep[e.g.][]{Ward-Thompson94, Ward-Thompson99, Shirley00}.  To study
the density profiles of the starless cores in our sample, we use the
mass and radius derived from the 1.1\,mm observations to predict the
3\,mm flux at the center of each core with the assumptions that the
density profile goes as $r^{-2}$, that the emissivity spectral index
is given by $\beta = 2$, and that the cores are isothermal.  We would
have been able to detect (at the 3$\sigma$ level) all 11 cores in our
sample if these assumptions were valid, but in fact we only detect two
starless cores.  As noted in Section \ref{MASS}, if $\beta=1$ then the
detection rate of cores, and any substructure in them, would be easier
than if $\beta=2$.  Because we expect starless cores to be colder in
their centers than at the edges the flux profile of a ``real''
starless core will be somewhat flatter than our isothermal model would
predict.  Nevertheless, we conclude that the density profiles for the
majority of the starless cores in our sample are flatter than an
$r^{-2}$ profile at their centers, in agreement with other surveys.

The low rate of 3\,mm continuum detections towards 11 of the brightest
\citep[$>$200 mJy beam$^{-1}$ at 1.1\,mm with Bolocam;][]{Enoch08} in
this study is in agreement with the results of \citet{Olmi05} who
detect only 1 out of 6 dense cores in Perseus.  The largest angular
scale to which the interferometer was sensitive and the median noise
in the maps in this study (30\arcsec--40\arcsec\ and 0.7 mJy
beam$^{-1}$) are similar to that in \citet[20\arcsec--40\arcsec\ and
1-1.5 mJy beam$^{-1}$;][]{Olmi05}.  There is no overlap in the sample
of sources mapped in this study and in \citet{Olmi05}.

There are 67 starless cores detected by \citet{Enoch08} in the Perseus
molecular cloud, of which we observed 11.  Because we selected our
sample based on their high peak fluxes, there is no reason to expect
that the other starless cores in Perseus would have a higher detection
rate at 3\,mm.  The low rate (0/11) of finding mutiple 3\,mm
components at 5\arcsec\ resolution within 1.1\,mm cores observed with
30\arcsec\ resolution is consistent with the 3\,mm continuum OVRO
observations of dense cores in Perseus carried out by \citet{Olmi05},
who found that none of the cores (out of six) were seen to be composed
of multiple components.  We argue that starless core mass functions
generated from single dish bolometer maps of Perseus
\citep[e.g.][]{Enoch08, Hatchell08} are not strongly biased by the
blending of compact, close cores into more massive single cores.

\section{Discussion} \label{DISCUSSION}

Given that many stars are found in multiple systems
\citep{Duquennoy91} and that many protostars are also multiples
\citep{Looney00, Haisch04}, it is worthwhile to examine when
fragmentation takes place.  If starless cores fragment before
formation of an embedded object, then we would expect to be able to
detect these sub-cores.  On the other hand, if fragmentation takes
place after the first protostar has formed, then a survey of starless
cores won't find any evidence of fragmentation.  In this study, we
find that none of the eleven starless cores show any evidence of
fragmentation.  If these cores will eventually fragment, then this
must happen during and/or after the collapse and formation of the
first protostar.  Late fragmentation on small scales, which is
consistent with the results of this survey, is predicted by theories
that create binaries from disk fragmentation \citep[e.g.][]{Bonnell94,
Kratter10}.

The observations described in this paper are in agreement with the
results of other recent surveys in Perseus. \citet{Kirk07} compared
the line profiles of C$^{18}$O (2-1) and N$_2$H$^+$(1-0) toward a
sample of 150 candidate dense cores, both prestellar and
protostellar. They found that the N$_2$H$^+$ lines, representing the
densest gas within the core, display nearly thermal line widths,
indicating little non-thermal activity at the core center. While the
C$^{18}$O gas shows significant non-thermal broadening, the velocity
centroids of the two species are very similar, indicating that the
connection between the cores and their environment is relatively calm.
Interestingly, such quiescence across scales appears hard to reconcile
with numerical simulations of turbulent clouds \citep{KirkH09} and may
indicate a need for magnetic pressure support within the dense gas.
Additionally, \citet{Jorgensen07} compared the location of
mid-infrared {\it Spitzer} sources and sub-millimeter cores in
Perseus, finding excellent agreement between the deeply embedded
protostars and their proto-stellar envelopes. Intriguingly, only 3 out
of 40 cores were found to harbour multiple protostars, at the spatial
resolution of {\it Spitzer} (MIPS 24\,\micron, 6\arcsec\ or
$\sim$1500\,AU). Thus, it appears that within Perseus there is little
evidence for fragmentation within cores, at least until the
protostellar stage, where the fragmentation length scale is
smaller. This may be due to the smooth and quiescent manner in which
the cores themselves appear to be forming.  

We only detected significant ($\ge$5$\sigma$) emission from two of the
eleven starless cores in our sample and we put tight upper limits on
the masses of multiple components in the other nine cores, from which
we conclude that there is no substructure on $\sim$5\arcsec\ scales
($\sim$1200\,AU) in the majority of starless cores.  A smooth column
density profile is predicted by many models of isolated star
formation, such as the inside-out collapse model \citep{Shu77} and
ambipolar diffusion \citep{Ciolek93}.  It is less clear if a smooth
column density profile is consistent with cores created through
turbulence, though some simulations show that turbulent molecular
clouds do form quiescent cores with smooth column density profiles
\citep{Nakamura05, Ballesteros03}.

\section{Summary} \label{SUMMARY}

We present a study of the 3\,mm continuum emission from a sample of 11
of the brightest (at 1.1\,mm) starless cores in the Perseus molecular
cloud.  The data were primarily taken with CARMA at $\sim$5\arcsec\
resolution, with followup observations of one core taken with the SZA
to improve the sensitivity to extended flux.

1) We detect two starless cores (Perbo45 and Perbo58), both of which
are seen to contain a single, main component.

2) Based on the low detection rate \citep[0/11 in this study and 0/6
  in][]{Olmi05} of multiple components within starless cores, we
  conclude that starless core mass functions derived from bolometer
  maps of the Perseus molecular cloud \citep[e.g.][]{Enoch08,
  Hatchell07} are unlikely to be heavily biased by the blending of
  smaller, nearby cores into single sources at the coarser resolutions
  of single-dish maps.

3) Based on the low detection rate of any components within dense
cores \citep[2/11 in this study and 1/6 in][]{Olmi05}, we conclude
that the majority of starless cores lack compact structures.  More
specifically, we find that density profiles in the inner portions of
starless cores of the form $r^{-2}$ are inconsistent with our
obervations for the majority of the starless cores in our survey.
This agrees with the results of previous surveys, which find that the
density profiles of starless cores are flatter than $r^{-2}$ at small
radii \citep[e.g.][]{Ward-Thompson94, Ward-Thompson99, Shirley00}.

\acknowledgments

We would like to thank John Carpenter, James Di Francesco, Brenda
Matthews and Sarah Sadavoy for helpful discussions.  SS acknowledges
support from a Plaskett Fellowship at the Herzberg Institute of
Astrophysics.  Support was provided to ME by NASA through the {\it
Spitzer Space Telescope} Fellowship Program.  DJ is supported by a
Natural Sciences and Engineering Research Council of Canada (NSERC)
Discovery Grant.  Support for DM was provided by NASA through Hubble
Fellowship grant HF-51259.01-A.  We thank the CARMA staff, students
and postdocs for their help in making these observations.  Support for
CARMA construction was derived from the Gordon and Betty Moore
Foundation, the Kenneth T. and Eileen L. Norris Foundation, the
Associates of the California Institute of Technology, the states of
California, Illinois and Maryland, and the National Science
Foundation.  Ongoing CARMA development and operations are supported by
the National Science Foundation under a cooperative agreement, and by
the CARMA partner universities.  The JCMT is operated by the Joint
Astronomy Centre on behalf of the Particle Physics and Astronomy
Research Council of the United Kingdom, the Netherlands Organisation
for Scientific Research, and the National Research Council of Canada.
The CSO is supported by the NSF fund under contract AST 02-29008.

{}

\begin{deluxetable}{lccccccc} 
\tablewidth{0pt}
\tabletypesize{\scriptsize}
\tablecaption{Core Properties from Bolocam 1\,mm Data \label{BOLOCAMTAB}}
\tablehead{
 \colhead{Name}       & 
 \colhead{RA}         & 
 \colhead{Dec}        & 
 \colhead{Peak Flux}  &
 \colhead{Total Flux} &
 \colhead{Total Mass} &
 \colhead{FWHM\tablenotemark{1}}       &
 \colhead{Density\tablenotemark{2}}     \\
 \colhead{}           &
 \colhead{(J2000)}    &
 \colhead{(J2000)}    &
 \colhead{(mJy/beam)} &
 \colhead{(Jy)}       &
 \colhead{(\solmass)} &
 \colhead{(\arcsec)}  &
 \colhead{(10$^5$ cm$^{-3}$)}}
\startdata
Perbo11  & 03:25:46.0 & +30:44:10 & 241 & 0.90 & 2.18 & 72.5 & 2.2 \\
Perbo13  & 03:25:48.8 & +30:42:24 & 407 & 0.47 & 1.14 & 38.2 & 3.9 \\
Perbo14  & 03:25:50.6 & +30:42:01 & 342 & 0.41 & 1.00 & 38.7 & 3.5 \\
Perbo44  & 03:29:04.5 & +31:18:42 & 274 & 0.72 & 1.73 & 47.0 & 2.7 \\
Perbo45  & 03:29:07.7 & +31:17:17 & 455 & 1.35 & 3.25 & 52.9 & 4.8 \\
Perbo50  & 03:29:14.5 & +31:20:30 & 313 & 1.31 & 3.15 & 68.0 & 3.2 \\
Perbo51  & 03:29:17.0 & +31:12:26 & 423 & 0.62 & 1.49 & 34.7 & 3.6 \\
Perbo58  & 03:29:25.7 & +31:28:16 & 273 & 0.33 & 0.78 & 25.9 & 1.8 \\
Perbo74  & 03:33:01.9 & +31:04:32 & 255 & 0.38 & 0.91 & 44.5 & 2.6 \\
Perbo105 & 03:43:57.8 & +32:04:06 & 283 & 0.61 & 1.47 & 46.6 & 2.8 \\
Perbo107 & 03:44:02.1 & +32:02:34 & 388 & 0.48 & 1.17 & 46.0 & 4.1
\enddata
\tablenotetext{1}{Derived from an elliptical Gaussian fit, deconvolved
by the 31\arcsec\ beam}
\tablenotetext{2}{Mean density calculated in a fixed linear aperture
of diameter 10$^4$\,AU}
\end{deluxetable}

\begin{deluxetable}{lcccccc} 
\tablewidth{0pt}
\tabletypesize{\scriptsize}
\tablecaption{3\,mm Observations \label{MMOBSTAB}}
\tablehead{
 \colhead{Name}       & 
 \colhead{CARMA D}    & 
 \colhead{CARMA E}    & 
 \colhead{SZA}        &
 \colhead{Beam Size\tablenotemark{1}}  &
 \colhead{LAS\tablenotemark{2}}        &
 \colhead{Noise}      \\
 \colhead{}           &
 \colhead{}           &
 \colhead{}           &
 \colhead{}           &
 \colhead{(\arcsec)}  &
 \colhead{(\arcsec)}  &
 \colhead{(mJy/beam)}}
\startdata
Perbo11  & Y & N & N & 4.5 & 30 & 0.5 \\
Perbo13  & Y & Y & N & 6.6 & 40 & 1.3 \\
Perbo14  & Y & Y & N & 6.2 & 40 & 0.9 \\
Perbo44  & Y & N & N & 4.2 & 30 & 0.6 \\
Perbo45  & Y & Y & N & 6.4 & 40 & 0.4 \\
Perbo50  & Y & N & N & 4.2 & 30 & 0.7 \\
Perbo51\tablenotemark{3}  & Y & Y & N & 5.4 & 40 & 2.8 \\
Perbo58  & Y & Y & Y & 5.6 & 80 & 0.3 \\
Perbo74  & Y & N & N & 4.6 & 30 & 0.3 \\
Perbo105 & Y & N & N & 4.3 & 30 & 0.9 \\
Perbo107 & Y & Y & N & 6.2 & 40 & 1.1
\enddata
\tablenotetext{1}{Geometric mean of the major and minor axes}
\tablenotetext{2}{Largest angular scale to which the map is sensitive}
\tablenotetext{3}{The noise in the map of Perbo51 is larger than the
other maps due to the bright source NGC 1333 IRAS4 located at the edge
of the CARMA map}
\end{deluxetable}

\begin{deluxetable}{lcccccccc} 
\tablewidth{0pt}
\tabletypesize{\scriptsize}
\tablecaption{Gaussian Fits to 3\,mm Observations \label{MMFITSTAB}}
\tablehead{
 \colhead{Name}          & 
 \colhead{RA offset\tablenotemark{1}}     & 
 \colhead{Dec offset\tablenotemark{1}}    & 
 \colhead{Peak Flux\tablenotemark{2}}     &
 \colhead{Total Flux\tablenotemark{2}}    &
 \colhead{Axes\tablenotemark{3}}          &
 \colhead{$\theta_{PA}$\tablenotemark{3}} & 
 \colhead{Mass\tablenotemark{4}}          &
 \colhead{density}                        \\
 \colhead{}              &
 \colhead{(\arcsec)}     &
 \colhead{(\arcsec)}     &
 \colhead{(mJy/beam)}    &
 \colhead{(mJy)}         &
 \colhead{(\arcsec)}     &
 \colhead{(degrees)}     &
 \colhead{(\solmass)}    &
 \colhead{(cm$^{-3}$)}}
\startdata
Perbo11     &                &                 &               &               &                &     & $<$0.11 \\
Perbo13     &                &                 &               &               &                &     & $<$0.29 \\
Perbo14     &                &                 &               &               &                &     & $<$0.20 \\
Perbo44     &                &                 &               &               &                &     & $<$0.14 \\
Perbo45\tablenotemark{5}     & $-10.3 \pm 0.5$ &   $8.8 \pm 0.7$ & $2.4 \pm 0.3$ & $11 \pm 0.5$ & $14 \times 9$ & -14 & 0.8 & $1.1 \times 10^7$ \\
Perbo50     &                &                 &               &               &                &     & $<$0.16 \\
Perbo51     &                &                 &               &               &                &     & $<$0.62 \\
Perbo58     & $-4.3 \pm 0.8$ &  $-1.1 \pm 0.9$ & $2.0 \pm 0.3$ & $33 \pm 1$ & $26 \times 18$ & 35  & 2.4 & $4.5 \times 10^6$ \\
Perbo74     &                &                 &               &               &                &     & $<$0.07 \\
Perbo105    &                &                 &               &               &                &     & $<$0.20 \\
Perbo107    &                &                 &               &               &                &     & $<$0.24
\enddata
\tablenotetext{1}{Offset measured from pointing center, given in Table \ref{BOLOCAMTAB}}
\tablenotetext{2}{Errors derived from the random noise in the maps and do not include absolute flux calibration uncertainties}
\tablenotetext{3}{Deconvolved major and minor axes and position angle}
\tablenotetext{4}{For non-detections, the 3$\sigma$ upper limit to a point source mass is given}
\tablenotetext{5}{Note that only CARMA D and E-array obserations have
been taken for this source, so we are likely to have resolved out a
larger fraction of the emission than for Perbo58}
\end{deluxetable}

\end{document}